# Higgs-Leggett mechanism for the elusive $6e$ superconductivity in Kagome material CsV$_3$Sb$_5$

## Author Information

Ling-Feng Zhang[1], Zhi Wang[2], and Xiao Hu[3]*

[1]Shanghai Key Laboratory of High Temperature Superconductors, Shanghai Frontiers Science Center of Quantum and Superconducting Matter States, Department of Physics, Shanghai University；Shanghai 200444, China.

[2]School of Physics, Sun Yet-sen University; Guangzhou 510275, China.

[3]International Center for Materials Nanoarchitectonics (WPI-MANA), National Institute for Materials Science; Tsukuba 305-0044, Japan.

*Corresponding author's Emails: Hu.Xiao@nims.go.jp

## Abstract

A recent Little-Parks experiment in Kagome-structured superconductor CsV$_3$Sb$_5$ demonstrated remarkable resistance oscillations with period $\phi_0/3 = hc/6e$. Here, we perform analysis based on a theory involving three $2e$ superconductivity (SC) order parameters associated with the three reciprocal lattice vectors which connect M points of the hexagonal Brillouin zone. In a ring geometry we unveil that, as a series of metastable states, phase of one SC order parameter winds $2\pi$ more or less than the other two ones around the ring, which yields local free energy minima at integer multiples of $\phi_0/3$. Intriguingly, the ground-state degeneracy associated with a $Z_2$ chirality is crucial, and the *Higgs-Leggett* mechanism stabilizes domain walls (DW) between chiral domains. At low temperatures DW are expelled from the system resulting in free energy minima only at integer multiples of $\phi_0$. Our theory explains successfully the $6e$ SC observed in experiments, which opens a door for approaching rich physics of Kagome superconductors.

## Introduction

Quantization of magnetic flux $\phi_0 = hc/2e$ is the hallmark of superconductivity (SC) as a global quantum state of electron pair condensation [1–6] (for a textbook see [7]). As shown first by Little and Parks [8], in a ring-shaped superconductor the SC transition temperature and resistance between two electrodes oscillate as functions of applied magnetic flux through the ring with period $\phi_0$. Very recently, in the new Kagome vanadium-based superconductors $CsV_3Sb_5$ [9–18] a Little-Parks experiment demonstrated clear resistance oscillations of period $\phi_0/3$, hinting existence of a fantastic six electrons ($6e$) SC state at intermediate temperatures slightly above the bulk transition point for vanishing zero-field resistance [19]. This remarkable observation raises several questions which may be intimately related with the fundamental understanding on SC [20]: Does the period $\phi_0/3$ of magneto resistance imply a condensate of six electrons ($6e$) different from the $2e$ Cooper pairs? How does the Kagome structure play the role in this phenomenon, and is it possible that Cooper pairs associated with the three important reciprocal lattice vectors of the Kagome structure interweave with each other in a unique way to give birth to the never-seen-before period $\phi_0/3$? After all, can and in what sense this phase be considered a SC state, and how does the ordinary $2e$ SC state re-assert itself at lower temperatures where zero-field resistance drops to zero?

Here, we propose a theory with three SC order parameters for Cooper pairs associated with the three reciprocal lattice vectors connecting M points of the hexagonal Brillouin zone of the material. Based on analysis of the free energy functional on a ring geometry mimicking the Little-Parks setup, we find that, in a series of metastable states, phase of one SC order parameter winds $2\pi$ more or less than the other two ones around the ring, which yields local free energy minima at integer multiples of $\phi_0/3$ starting from zero applied magnetic flux, responsible for the magneto resistance oscillations with period of $\phi_0/3$ observed in the recent Little-Parks experiment for CsV$_3$Sb$_5$ [19]. As the mechanism of this remarkable phenomenon, the existence of degeneracy associated with a $Z_2$ chirality in the ground state is crucial, and in the competition between the *Higgs mode* [21,22] and the *Higgs-Leggett mode* [23,24] the latter wins and stabilizes domain walls (DW) of $Z_2$ chirality domains defined by phases of the three SC order parameters [24–27]. Expelling these DW from the system at low temperatures brings the system to its ground state where the phases of three SC order parameters are locked to each other and behave in the same way as a single SC order parameter, yielding free energy minima only at integer multiples of $\phi_0$ and conventional Little-Parks signals. Therefore, we explain successfully the main results of the recent Little-Parks experiment [19], namely the 6$e$ SC state observed at intermediate temperatures slightly above the bulk SC transition point and the re-assertion of 2$e$ SC at low temperatures. It is also revealed that a similar 4$e$ SC state associated with half magnetic flux quantization $\phi_0/2$ is possible which depends on the detailed Josephson-type coupling. The present theory is expected to provide a reasonable starting point for upcoming exploration on the rich physics hosted by the Kagome vanadium-based superconductors.

## Results

*Free energy functional.* We consider the free energy functional of three SC order parameters for Cooper pairs under an external magnetic field. For the cause for simplicity, it is taken in the Ginzburg-Landau (GL) form including only quadratic and quartic terms of SC order parameters (for details see Supplementary Materials)

$$G = \sum_{j=1,2,3} \left[ a_j |\psi_j|^2 + \frac{b_j}{2} |\psi_j|^4 + \frac{1}{2m_j} \left| \left( \frac{\hbar}{i} \boldsymbol{\nabla} - \frac{2e}{c} \boldsymbol{A} \right) \psi_j \right|^2 \right]$$

$$- \sum_{j,k=1,2,3; j<k} \gamma_{jk} (\psi_j^* \psi_k + \text{c.c.}) + \frac{1}{8\pi} (\boldsymbol{\nabla} \times \boldsymbol{A} - \boldsymbol{H})^2 \quad (1)$$

where due to the Kagome lattice symmetry $a_j = a$, $b_j = b$, $m_j = m$ and $\gamma_{jk} = \gamma$ for $j, k = 1, 2, 3$; $b, m > 0$ and $\gamma < 0$. The mean-field SC transition point $T_{\text{mf}}$ of the system including the inter-component Josephson-type couplings is given by $a + \gamma = 0$ [24], below which the order parameters take finite amplitude $|\psi_j| = \sqrt{-(a+\gamma)/b}$, and the two coherence lengths and the penetration depth of magnetic field diverge. In the present work, the three SC order parameters are related to the three reciprocal lattice vectors of the Kagome lattice. It is worthy noticing that the bilinear Josephson-type couplings including their sign are crucial for the $Z_2$ chiral SC state discussed in systems with three and more SC order parameters [24–27] (for more works see the review article [28]). The second-order Josephson-type coupling discussed in the context of pair density wave (PDW) states [29–31] will be considered later.

*Analysis for ring geometry.* In order to simulate the Little-Parks experiment we consider a SC ring with magnetic flux penetrating as shown schematically in Fig. 1(a). For clarity we mainly consider the small ring width as compared with the ring radius. Performing numerical calculations we look for stable and/or metastable states of the system when the applied magnetic flux is swept below the mean-field transition temperature $T_{\text{mf}}$ (with $a + \gamma < 0$) (for details see Methods). As displayed in Fig. 1(b), there is a series of ground states with their free energy minima at integer multiples of flux quantization $\phi_0$, such as $[000]_g$, $[111]_g$, $[222]_g$ and $[333]_g$, where the number $L = 0, 1, 2, \ldots\ldots$ refers to the winding numbers of the three SC order parameters around the SC ring. Namely the three SC order parameters wind simultaneously in these ground states although the three phases deviate from each other by $2\pi/3$ due to the repulsive Josephson-type coupling $\gamma < 0$ in Eq. (1). In these states the system responds to the external magnetic field in the same way as single-component superconductors [8]; especially the magneto resistance oscillates with applied magnetic flux in the period $\phi_0$ as seen in the recent Little-parks experiment at low temperatures [19].

Remarkably, we find another series of *metastable* states at free energy higher than the ground states which exhibit their free energy minima at integer multiples of $\phi_0/3$ starting from the zero applied magnetic flux. Distinguished from the ground states, one phase of the three SC order parameters winds $2\pi$ more than the other two in states [100], [211], [322] etc., or less than the other two in states [011], [122], [233] etc., which intervene each other and yield the free energy minima separated from each other by $\phi_0/3$. We believe these metastable states are responsible for the $\phi_0/3$ magneto resistance oscillations in the recent Little-Parks experiment at temperatures ranging from 2.9K to 2.4K [19].

The phase winding in these metastable states is much richer as compared with the ground states. In state [100] depicted in Fig. 2(a) and 2(b), since the SC order parameter $\psi_1$ acquires $2\pi$ phase around the ring, its phase changes faster than the other two SC order parameters, which renders inevitably two DW: one at the north pole of the ring where the phases of $\psi_1$ and $\psi_2$ across each other, the other one at the south pole of the ring where the phases of $\psi_1$ and $\psi_3$ across each other as shown in Fig. 2(b). As addressed in previous works on SC states with three order parameters and $\gamma < 0$ (see Eq. (1)), a $Z_2$ chirality can be defined based on the relative phase differences when the bilinear Josephson-type coupling takes $\gamma < 0$ in Eq. (1), where the time reversal symmetry is broken spontaneously [24–28]. It is clear that in state [100] the two halves of the SC ring correspond to two domains with opposite chirality as depicted by the red and blue colors in the circle at the ring center in Fig. 2(b). The DW are against the Josephson couplings, which raises the free energy of state [100] above the ground state. The system compromises the energy cost by tuning amplitudes of SC order parameters according to the total free energy expression as can be seen in Fig. 2(b) for state [100]. When the winding numbers increase, the three phases interweave in magnificent patterns forming beautiful phase kaleidoscopes as displayed in Fig. 2(b).

Metastable states should also take place at integer multiples of $\phi_0$ in order to complete the $\phi_0/3$ series. Comparing metastable states [000], [111], [222] and [333] with same winding numbers in the three SC order parameters in Fig. 2(b) with their counterparts in the ground state $[000]_g$, $[111]_g$, $[222]_g$ and $[333]_g$ in Fig. 2(c), it is clear that in the metastable states there are two DW between same two SC order parameters, such as $\psi_1$ and $\psi_2$, whereas in other metastable states shown in Fig. 2 the two DW involve three SC order parameters and there is no DW in ground states [32]. There are other possible metastable states including four or more DW, which take higher free energies and are thus expected at higher temperatures. Nevertheless, their free energy minima always take place at integer multiples of $\phi_0/3$, same as the metastable states with 2 DW displayed in Fig. 1(b) and Fig. 2.

Starting from one of the metastable states such as state [100] and increasing the applied magnetic flux, the system remains to the same winding configuration up to a magnetic flux $\phi/\phi_0=0.885$, which is much higher than $\phi/\phi_0=2/3$ where the free energy minimum of state [011] locates. As the price, the system acquires a large free energy which is induced by the large suppression in SC order parameters $\psi_2$ and $\psi_3$ as depicted in Fig. 2(d). Then, upon a tiny increase of magnetic flux to $\phi/\phi_0=0.89$ state [100] loses its metastability and the system jumps to state [111] with a smaller free energy accompanied by the recovery of the amplitudes of $\psi_2$ and $\psi_3$. Obviously, this is nothing but penetration of two vortices into the SC ring, each carried by $\psi_2$ and $\psi_3$. Similarly, around $\phi/\phi_0=2.215$ two vortices penetrate into the SC ring, both carried by $\psi_1$.

Why can states with winding numbers distinct in the three SC order parameters be metastable? First of all, there is a double degeneracy in the ground state associated with a $Z_2$ chirality defined by the relative phases of the three SC order parameters in the present system with repulsive bilinear Josephson-type coupling between three SC order parameter ($\gamma < 0$ in Eq. (1)). In addition, there are two diverging coherence lengths [24]: $\xi_H = \hbar/\sqrt{-2(a+\gamma)m}$ associated with the *Higgs mode*, where only the amplitudes of SC order parameters are involved (see Fig. 3(a)) which is essentially same as single-component SC [21,22]; $\xi_{HL} = \hbar/\sqrt{-(a+\gamma)m}$ associated with a *Higgs-Leggett* *mode*, which involves both SC amplitudes and phases (see Fig. 3(b)) [23,24]. As indicating by $\xi_{HL} > \xi_H$, the free energy cost of distortions in SC order parameters induced by the Higgs-Leggett mode is smaller than that induced by the Higgs mode (the zero-energy Nambu-Goldstone mode corresponds to an infinite coherence length). With the Higgs-Leggett mechanism, the system responds to the applied magnetic flux by enhancing amplitude in one SC order parameter and squeezing phase difference between the two remaining SC order parameters as schematically shown in Fig. 3(b) [24] (see also Fig. 2(d)); when the phase difference shrinks to zero and then changes sign, a DW between $Z_2$ chirality domains appears. This renders states with winding numbers distinct in the three SC order parameters, and thus local free energy minima at integer multiples of $\phi_0/3$, as revealed above (see Fig. 2). In a stark contrast, in the Higgs mechanism, all three SC order parameters carry the same winding number and are suppressed simultaneously (see Fig. 3(a)); in order to accommodate more vortices into the SC ring, all the three SC order parameters have to be suppressed to zero, which costs a large free energy and thus is unfavorable.

***Second-order Josephson-type coupling.*** Instead of the first-order Josephson-type coupling, one can consider the second-order, bi-quadratic ones such as $-(\eta/2)\psi_j^{*2}\psi_k^2$ which were discussed in the context of pair density wave (PDW) [29–31]. Here, we concentrate on $\eta < 0$ and for stability $b + \eta > 0$ is presumed. In addition to the $Z_2$ chirality, the ground state gains additional degeneracy as shown in Fig. 4(a). Nevertheless, as displayed in Fig. 4(b) and (c), we find that free energy minima at integer multiples of $\phi_0/3$ appear in the metastable states, where phase configurations with one SC order parameter winding more or less than the other two ones accompanied by DW separating $Z_2$ chiral domains are stabilized by the Higgs-Leggett mechanism, similarly to the case of the first-order Josephson-type coupling. Due to the additional degeneracy in the ground state, a new $\pi$-phase kink $\Pi_{jk}$ appears where SC order parameters $\psi_j$ and $\psi_k$ become out-of-phase, as seen in Fig. 4(c).

However, the fractional magnetic flux quantization is not influenced directly since they do not change the $Z_2$ chirality.

## Discussions

*Mean-field and genuine SC phase transitions.* In the present work, the fractional flux quantization $\phi_0/3$ has been clarified below the transition point $T_{\mathrm{mf}}$ determined by the GL free energy functional. While these states with phase windings distinct in the three SC order parameters accompanied by DW of opposite $Z_2$ chirality are metastable in the GL theory, they are stabilized thermodynamically when thermal fluctuations are addressed appropriately. The transition point $T_{\mathrm{mf}}$ is around 3K for the CsV$_3$Sb$_5$ sample in the Little-Parks experiment (see Fig. 1c and Fig. 2i in Ref. [19]). It is anticipated that at a lower, genuine transition point $T_c$ these DW should be expelled completely from the system, where magnetic flux quantization takes place only at integer multiples of $\phi_0$ as revealed in the present work, which renders the vanishing bulk zero-field resistance simultaneously. From Fig. 1c and Fig. 2i in Ref. [19], it is estimated $T_c \approx 1K$. Therefore, the fractional magnetic flux quantization is characteristic of an intermediate SC phase in the Kagome vanadium-based superconductors, which is reminiscent of the picture proposed in the previous works [29–31] where melting of the PDW lattice results in a phase of uniform order of $6e$ SC condensate, which is secondary before melting (see also [33,34]).

*About half flux quantization $\phi_0/2$.* The present theory involving three SC order parameters cannot capture the half flux quantization $\phi_0/2$. Then, we have performed the similar analysis for a system of two SC order parameters with the first-order Josephson-type coupling by omitting one SC order parameter in Eq. (1). We can only find states with the same winding number in the two SC order parameters where the free energy minima appear at the integer multiples of $\phi_0$, whereas configurations with winding numbers distinct in the two SC order parameters are *unstable*. It is worthy noticing that, with two SC order parameters coupled by the second-order Josephson-type interactions, a $Z_2$ chirality appears and we find metastable states with free energy minima at integer multiples of $\phi_0/2$. This is considered to correspond to the magneto resistance oscillations with period of half magnetic flux quantization, and thus $4e$ superconductivity, observed in the recent Little-Parks experiment [19].

*Other possible novel phenomena.* For the chiral $6e$ SC state implied by the Little-Parks experiment [19] novel SC phenomena are expected in addition to the one discussed in the present work. For a SC ring with sufficient width where supercurrent is suppressed to zero due to Meissner effect at the center part of superconductor, the magnetic flux trapped by the SC ring should be quantized into integer multiples of $\phi_0/3$ resulting in plateaus in magnetization curve (see also [32]). In a narrow constriction between two bulk $CsV_3Sb_5$ crystals, the critical Josephson current is suppressed significantly (to zero in theory) when the two bulks take opposite chirality, as compared to the case where a same chirality occupies the two bulks [24]. An unconventional intermediate SC state characterized by clustering vortices may also be possible [35].

## Methods

*Numerical simulations.* In order to capture the period-$\phi_0/3$ magneto resistance observed in the Kagome vanadium-based superconductor, we consider the GL free energy functional with three SC order parameters $\psi_j$ ($j = 1,2,3$) associated with the three reciprocal lattice vectors connecting M points of the hexagonal Brillouin zone of the material as given in Eq. (1). Performing the variational analysis with respect to $\psi_j^*$, we obtain the GL equations:

$$a\psi_j + b|\psi_j|^2\psi_j + \frac{1}{2m}\left(\frac{\hbar}{i}\nabla - \frac{2e}{c}\boldsymbol{A}\right)^2\psi_j - \sum_{k=1,2,3;k\neq j}\gamma\psi_k = 0 \quad (2)$$

with $j = 1,2,3$. For numerical calculations, we use dimensionless quantities: $a$ and $\gamma$ in units of $a_0$, length in units of $\xi_1 = \hbar/\sqrt{-2ma_0}$, order parameter $\psi_j$ in units of $\psi_0 = \sqrt{-a_0/b}$, $\boldsymbol{A}$ in units of $\hbar c/2e\xi_1$, and free energy in units of $G_0=a_0^2/b$, where $a_0$ is a typical energy. In this work, we take $a = -0.1$ and $\gamma = -0.24$ referring to a temperature below the transition point $T_{\text{mf}}$, except otherwise noticed. For case of the second-order Josephson-type coupling, we replace the term $\gamma\psi_k$ in Eq. (2) with $\eta\psi_k^2\psi_j^*$, where $\eta = -0.15$ is taken in unit of $b$ for numerical calculations.

The simulation is implemented using finite difference method [36, 37] for ring geometry of the sample with polar coordinate. For rings with small width as compared to the penetration depth, the demagnetization effect can be neglected. The gauge $\nabla \cdot \boldsymbol{A} = 0$ is taken with $\boldsymbol{A} = \boldsymbol{e_\varphi} Hr/2$ for the uniform magnetic field $H$. The boundary conditions for $\psi_j$ correspond to zero total current density normal to the sample surface.

Starting from suitable initial configurations of SC order parameters, we obtain ground state solutions and metastable solutions by the iterative relaxation method. To find the metastable states with two DW carrying integer multiples of $\phi_0/3$, we should put initial guesses within the attractive basins of the final solutions. As an example, in order to obtain state [100] which carries magnetic flux of $\phi_0/3$ we set $\psi_1(\varphi) = \sqrt{n_0}e^{i\varphi}$, $\psi_2(\varphi) = \sqrt{n_0}e^{i2\pi/3}$ and $\psi_3(\varphi) = \sqrt{n_0}e^{i4\pi/3}$ where $n_0$ is the ground state amplitude in absence of the applied magnetic flux. Two DW separating two different chiral domains are formed after a few iteration steps from the initial guess during simulations. The algorithm is iterated until changes in SC order parameters between two steps are less than $10^{-6}$.

Next, we sweep slightly up/down the applied magnetic flux and recalculate the distribution of amplitudes and phases of SC order parameters with those of the previous solution as the initial guess, and reach the new solution. Repeating this process gives one of the parabolic curves in the diagram of free energy functional. Although they are not the ground states, we find that the solutions with two DW and winding numbers distinct in the three SC order parameters are stable against small perturbations. As the applied magnetic flux deviates largely from the value where a local minimum of free energy locates, the solution jumps discontinuously to a new one with a different set of winding numbers. This results in the series of metastable states with local free energy minima at integer multiples of $\phi_0/3$.

**Acknowledgments.** X. Hu acknowledges J. Wang and F.-C. Zhang for communications and discussions.

**Funding.** L.-F. Zhang is supported by the National Natural Science Foundation of China (Grant No. 62171267), in part by the National Key Research and Development Program of China (No. 2022YFE03150200), the Strategic Priority Research Program of the Chinese Academy of Sciences (No. XDB25000000) and the National Natural Science Foundation (No. 52172271). Z. Wang is supported by the National Natural Science Foundation of China (Grant No. 12174453). X. Hu is supported by CREST, JST (Core Research for Evolutionary Science and Technology, Japan Science and Technology Agency) (Grant No. JPMJCR18T4).

**Author contributions.** X.H. conceived the idea, supervised the project, and wrote the manuscript. L.F.Z. performed the numerical calculations. Z.W. joined discussions. All authors fully contribute to the research.

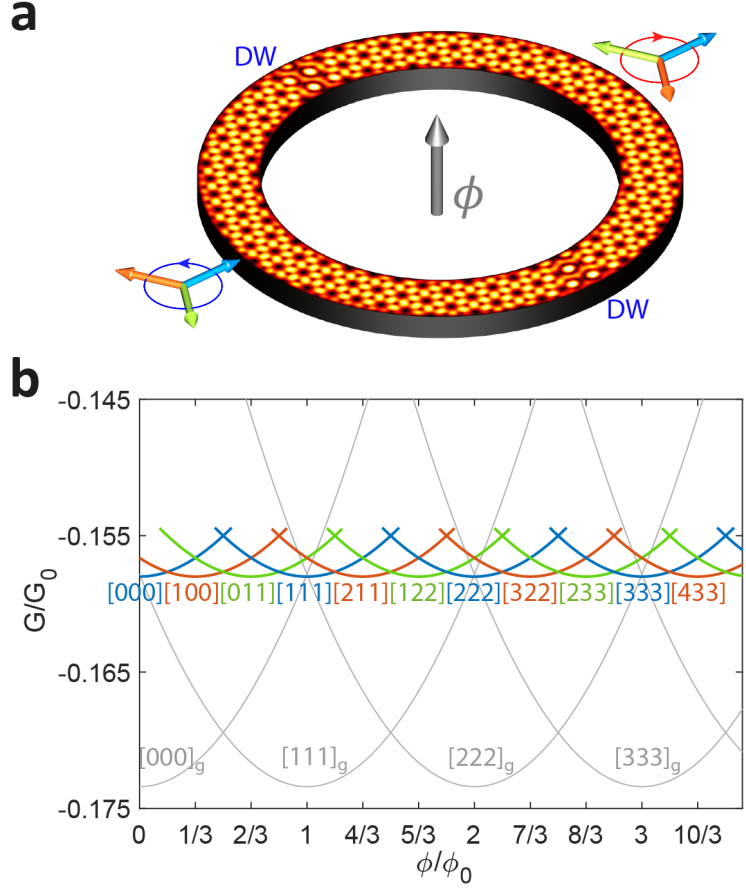

**Fig. 1. Free energy minima at integer multiples of one-third flux quantization $\phi_0/3$ in the Little-Parks setup for the Kagome-structured superconductor CsV$_3$Sb$_5$.** (a) Schematic setup for the Little-Parks experiment where $Z_2$ chirality domains defined by phases of three SC order parameters are separated by DW, which induce distortions in the real-space PDW lattice shown schematically by color. (b) Free energy as a function of applied magnetic flux penetrating through the superconductivity ring. In the ground states, such as $[000]_g$, $[111]_g$, $[222]_g$ and $[333]_g$, the three SC order parameters wind in the same way around the ring, which yields free energy minima at the integer multiples of $\phi_0$. In the metastable states one phase of the three SC component winds one more time than the other two in states such as [100], [211], [322] etc., or one less time than the other two in states such as [011], [122], [233] etc., which intervenes each other and yields the free energy minima separated from each other by $\phi_0/3$. The metastable states are triply degenerate.

Dimensionless parameters are taken as $a = -0.1$ and $\gamma = -0.24$. The radius of the ring is taken as $R = 8$.

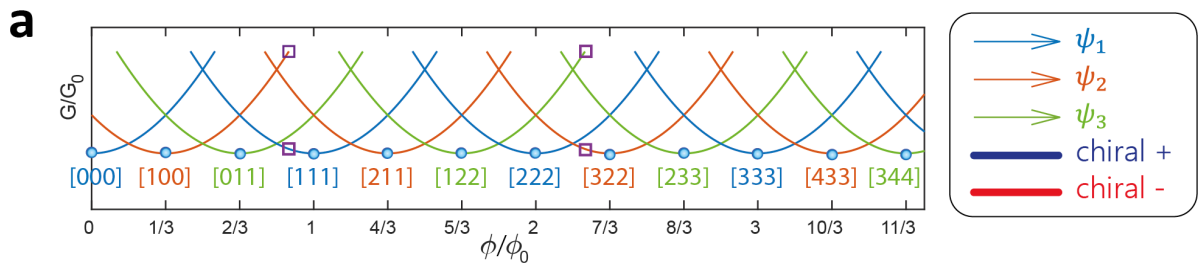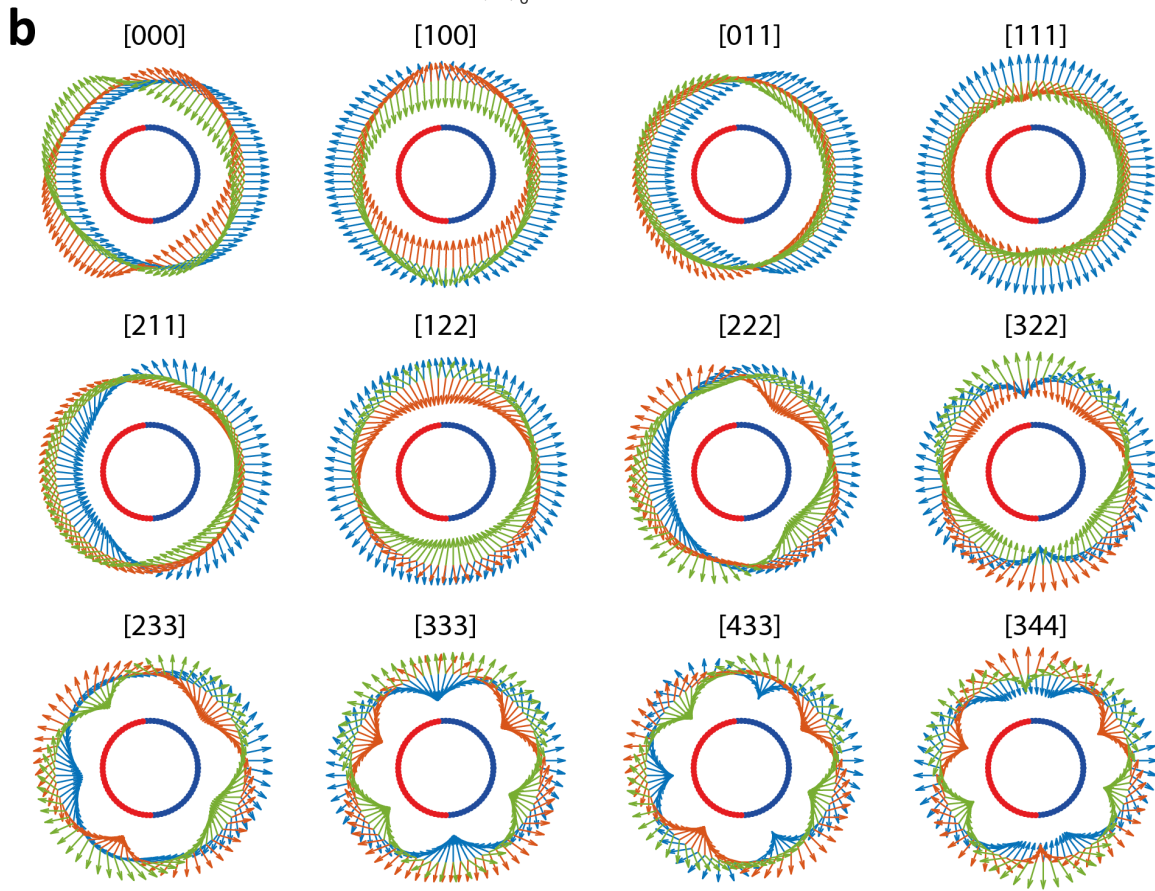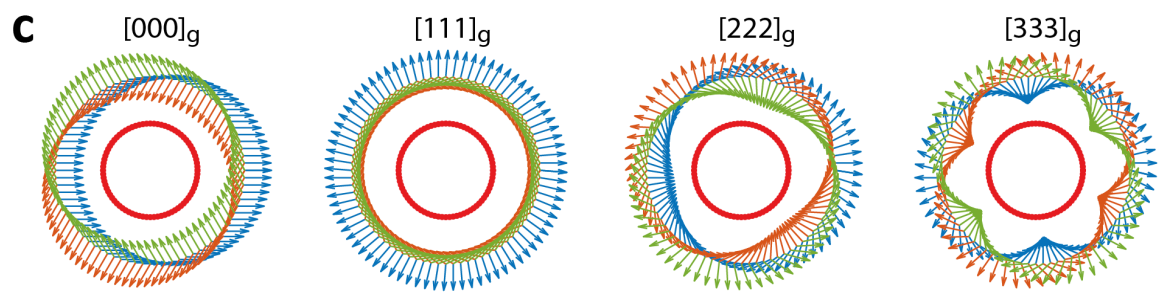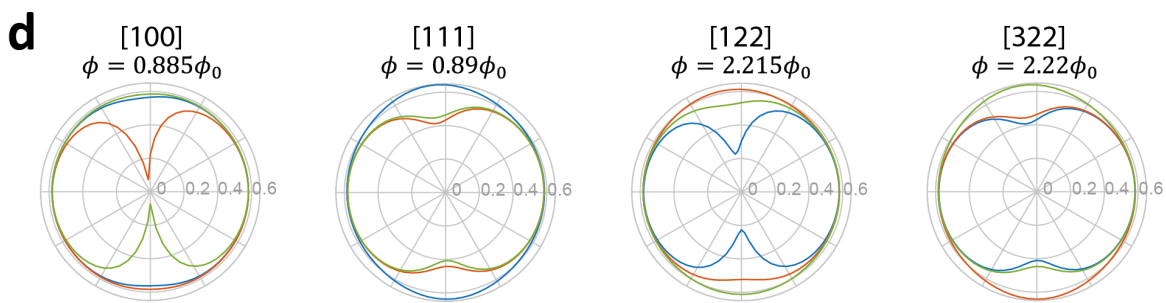

**Fig. 2. Kaleidoscope of phases winding for the three SC order parameters in the Little-Parks setup at integer multiples of $\phi_0/3$.** (a) Schematic diagram of free energy for the metastable states with circles and squares denoting the values of applied magnetic flux for which the detailed wavefunctions are displayed in (b) and (d), respectively. (b) Distributions of phases of the three SC order parameters around the SC ring in the metastable states at the applied magnetic flux marked by circles in (a) where, for example, [211] refers to the state with phase winding 2, 1 and 1 in the SC order parameter $\psi_1$, $\psi_2$ and $\psi_3$, respectively. (c) Same for (b) except for the ground states at applied magnetic flux of integer multiples of $\phi_0$. (d) Distributions of the amplitudes of the three SC order parameters around the SC ring in the metastable states at the applied magnetic flux marked by squares in (a), where the winding numbers jumps corresponding to penetration of vortices into the SC ring in one or two out of three SC order parameters. Parameters are taken same as Fig. 1.

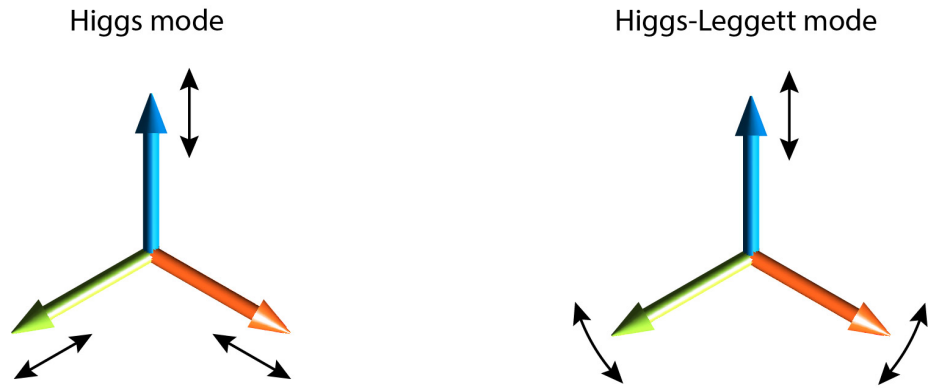

**Fig. 3. Possible modes in systems with three SC order parameters and repulsive first-order (bilinear) Josephson-type couplings.** (a) Higgs mode where only variations in amplitudes of the three SC order parameters are involved, which is essentially same for SC state with a single SC order parameter. (b) Higgs-Leggett mode where variation in amplitude of one SC order parameter is accompanied by variation of phase difference between the two remaining SC order parameters. The three colored arrows indicate the three complex SC order parameters associated with a $Z_2$ chiral state induced by the repulsive bilinear Josephson-type coupling.

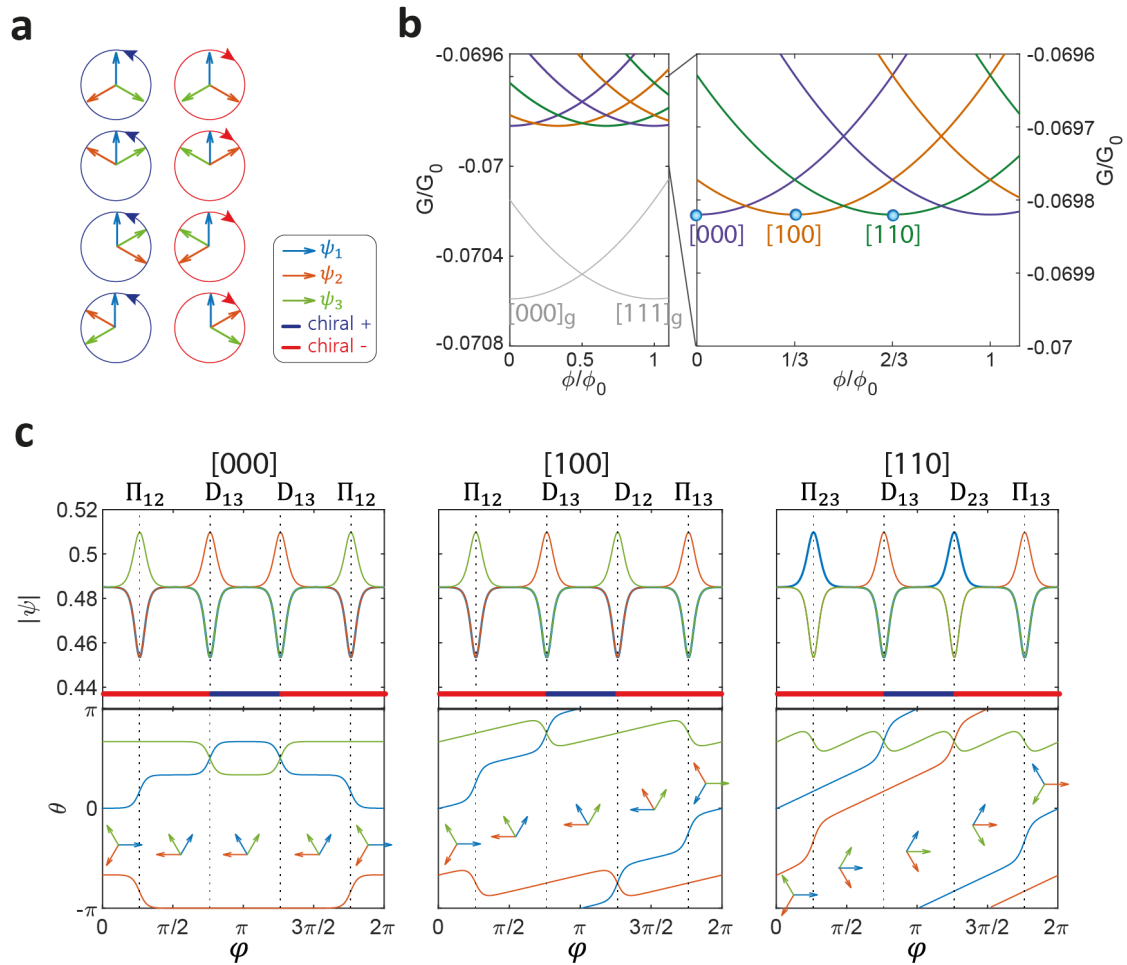

**Fig. 4. Free energy minima at integer multiples of one-third flux quantization $\phi_0/3$ and the corresponding metastable states for the second-order Josephson-type couplings.** (a) Degenerate ground states for the second-order Josephson-type couplings with $Z_2$ chirality, where the phase difference between one order parameter and the other two is $\pm 2\pi/3$ or $\pm \pi/3$. (b) Free energy as a function of applied magnetic flux through the ring for the ground (metastable) states with minima at integer multiples of $\phi_0$ ($\phi_0/3$). (c) Distributions of amplitudes and phases of the three SC order parameters around the SC ring in the metastable states at the applied magnetic flux marked by circles in (b). A new $\pi$-phase kink $\Pi_{jk}$ is observed where the SC order parameters $\psi_j$ and $\psi_k$ become out-of-phase, which does not change the $Z_2$ chirality. Domains with opposite $Z_2$ chirality as depicted by the red and blue colors are separated by two DW denoted by $D_{jk}$, and the insets show phase

configurations in individual domains. Dimensionless parameters are $a = -0.2$, $\eta = -0.15$ and $R = 40$.